\documentclass[12pt]{article}

\usepackage[margin=1in]{geometry}
\usepackage[utf8]{inputenc}
\usepackage[T1]{fontenc}
\usepackage{hyperref}
\usepackage{url}
\usepackage{booktabs}
\usepackage{tabularx}
\usepackage{array}
\usepackage{makecell}
\usepackage{float}
\usepackage{subcaption}
\usepackage{graphicx}
\usepackage{amsmath}
\usepackage{natbib}
\usepackage{listings}
\usepackage{microtype}
\usepackage{authblk}

\hypersetup{
    colorlinks=true,
    linkcolor=blue,
    citecolor=blue,
    urlcolor=blue,
    pdftitle={Adaptive Virtual Reality Museum: A Closed-Loop Framework for Engagement-Aware Cultural Heritage},
    pdfauthor={Joseph Damouni, Wadia Tanus, Naomi Unkelos-Shpigel},
}

\title{Adaptive Virtual Reality Museum: A Closed-Loop Framework\\
for Engagement-Aware Cultural Heritage}

\author[1]{Joseph Damouni}
\author[1]{Wadia Tanus}
\author[1]{Naomi Unkelos-Shpigel\thanks{Corresponding author: \texttt{naomius@braude.ac.il}}}

\affil[1]{Braude College of Engineering, Karmiel, Israel\\
  \texttt{\{joseph.damouny, wtanoos905\}@gmail.com}}

\date{}

\begin{document}

\maketitle

\begin{abstract}
Static information presentation in VR cultural heritage often causes cognitive overload or under-stimulation. We introduce a closed-loop adaptive interface that tailors content depth to real-time visitor behavior through implicit multimodal sensing. Our approach continuously monitors gaze dwell, head kinematics, and locomotion to infer engagement via a transparent rule-based classifier, which drives a Large Language Model to dynamically modulate explanation complexity without interrupting exploration.

We implemented a proof-of-concept in the Berat Ethnographic Museum and conducted a preliminary evaluation ($N=16$) comparing adaptive versus static content. Results indicate that adaptive participants demonstrated 2--3$\times$ increases in reading engagement and exploration time while maintaining high usability (SUS = 84.3). Technical validation confirmed sub-millisecond engagement inference latency on consumer VR hardware.

These preliminary findings warrant larger-scale investigation and raise questions about engagement validation, AI transparency, and generative models in heritage contexts. We present this work-in-progress to spark discussion about implicit AI-driven adaptation in immersive cultural experiences.
\end{abstract}

\noindent\textbf{Keywords:} Virtual Reality \quad$\cdot$\quad Cultural Heritage \quad$\cdot$\quad Adaptive Interfaces \quad$\cdot$\quad User Engagement

\bigskip

\section{Introduction}

In recent years, virtual reality (VR) has emerged as a powerful medium for cultural heritage presentation, enabling immersive reconstructions of historical spaces that are otherwise geographically inaccessible, physically fragile, or no longer extant \cite{marto2024}. By supporting spatial exploration and embodied interaction, VR museums promise richer engagement than traditional screen-based digital exhibits \cite{konstantakis2023}.

Despite these advantages, most existing VR museum systems adopt a static content delivery paradigm inherited from physical exhibitions. Exhibit descriptions are typically authored in advance and presented identically to all visitors, regardless of differences in attention span or moment-to-moment engagement \cite{khan2025museum}. Consequently, users frequently experience cognitive misalignment: cognitive overload when complex explanations are presented during rapid movement or under-stimulation when brief descriptions fail to satisfy focused attention \cite{Mahsa2025}.

While modern standalone VR headsets provide continuous streams of behavioral data (e.g., head motion, gaze direction, locomotion patterns, and hand movement) that offer the potential to infer attention without interrupting the user experience, in most VR museum applications these signals remain underutilized \cite{clay2023}. Recent advances in large language models have further expanded the design space by enabling generation of textual explanations with controllable length and complexity, though their integration into latency-sensitive interactive systems remains underexplored.

In this paper, we present a multimodal closed-loop adaptive framework for VR museum experiences that personalizes exhibit explanations in real-time based on inferred visitor engagement. Our approach monitors implicit behavioral signals available on a consumer VR headset, including gaze dwell, head kinematics, locomotion velocity, and hand proximity. These signals are processed by a transparent, rule-based engagement classifier that estimates the user's current engagement state without requiring calibration or explicit input. The inferred engagement level drives an adaptive content mechanism that leverages a large language model to modulate the depth of textual explanations \cite{steiner2026ar}. This work investigates the question:
\begin{quote}
\textit{Does engagement-driven content adaptation in a VR museum improve the user experience compared to a static VR museum?}
\end{quote}

We focus on two specific sub-questions:
\begin{itemize}
    \item \textbf{RQ1:} To what extent can implicit behavioral signals support real-time attention-aware adaptation?
    \item \textbf{RQ2:} Compared to a static control, how does adaptation influence exploration behavior, content consumption, and the motivation to engage with exhibits?
\end{itemize}

We implemented the system within a high-fidelity virtual reconstruction of the Berat Ethnographic Museum \cite{beratmuseum} and conducted a between-subjects user study ($N=16$) to evaluate these questions.

In summary, this paper makes the following contributions:
\begin{enumerate}
    \item A novel multimodal closed-loop framework that fuses implicit behavioral signals to infer user engagement in real-time.
    \item A proof-of-concept implementation showing a content adaptation pipeline that integrates Large Language Models (LLMs) into VR to dynamically adjust text complexity based on user attention.
    \item Initial empirical insights into the impact of implicit, pacing-aware personalization on behavioral engagement and content consumption in virtual museums.
\end{enumerate}

\section{Background and Related Work}

\subsection{Adaptive Interfaces in VR}

Adaptive interfaces are systems that modify their presentation, navigation, or interaction logic in response to user signals. In the context of VR this often involves tailoring the environment to reduce cybersickness or manage cognitive load. Prior research in Human-Computer Interaction (HCI) has investigated adaptation in training, rehabilitation, and gaming \cite{Yildirim17012025}. Within cultural heritage applications, personalization has primarily focused on pre-visit profiling or manually selected content paths. However, such approaches assume stable user intent \cite{ivanov2024advanced}. Design thinking driven AR heritage systems have also been used to enhance visitor engagement in cultural and memorial contexts \cite{unkelos2024heights}.

Recent research reports that static content presentation in immersive media can induce elevated extraneous cognitive load, suggesting that explicit selection mechanisms alone are insufficient to address dynamic attentional demands \cite{lee2025reading}. While recent work has explored multimodal intent modeling \cite{han2025} and LLM integration in XR \cite{tang2025}, our work differs by specifically targeting the pacing of information delivery using a closed-loop system, rather than general intent detection. Furthermore, unlike prior personalization approaches that rely on pre-visit profiling or explicit narrative selection \cite{lau2025,kuflik2011}, our framework adapts content depth solely through implicit, real-time behavioral cues, reducing reliance on manual calibration or menu-based interactions.

\subsection{Implicit Behavioral Signals}

User engagement is commonly described as a multidimensional construct encompassing attention, interest, and involvement. In the absence of explicit feedback (like buttons or ratings), systems must rely on implicit signals to estimate this state. Head kinematics provide a reliable proxy for attentional stability in VR \cite{inproceedings}, showing that low head angular velocity is strongly associated with sustained visual attention. Regarding gaze-based interaction, studies indicate that dwell durations on the order of one second are sufficient to distinguish intentional focus from transient glances in gaze-driven interaction pipelines \cite{10.1145/3607822.3614513}. Similarly, locomotion behavior has been shown to influence cognitive availability; specifically, increased movement is associated with reduced attentional resources during VR walking tasks \cite{10.1145/3343036.3343119}.

\subsection{Generative Content Adaptation}

Generative content adaptation refers to the use of AI to dynamically create or modify media at runtime. Recent advances in Large Language Models (LLMs) have enabled the generation of fluent, context-aware textual explanations. However, integrating LLMs into immersive VR environments introduces challenges related to latency and predictability. Research examining text consumption in VR indicates that effective reading requires perceptual stability to maintain readability. Due to these latency and reliability constraints, many prior applications of LLMs in immersive cultural heritage settings have prioritized offline generation or pre-scripted retrieval over fully real-time adaptation \cite{cho2026mobile}.

\section{Method}

\subsection{Engagement Modeling}

\begin{figure}[htbp]
  \centering
  \includegraphics[width=0.85\textwidth, trim=60 40 60 40, clip]{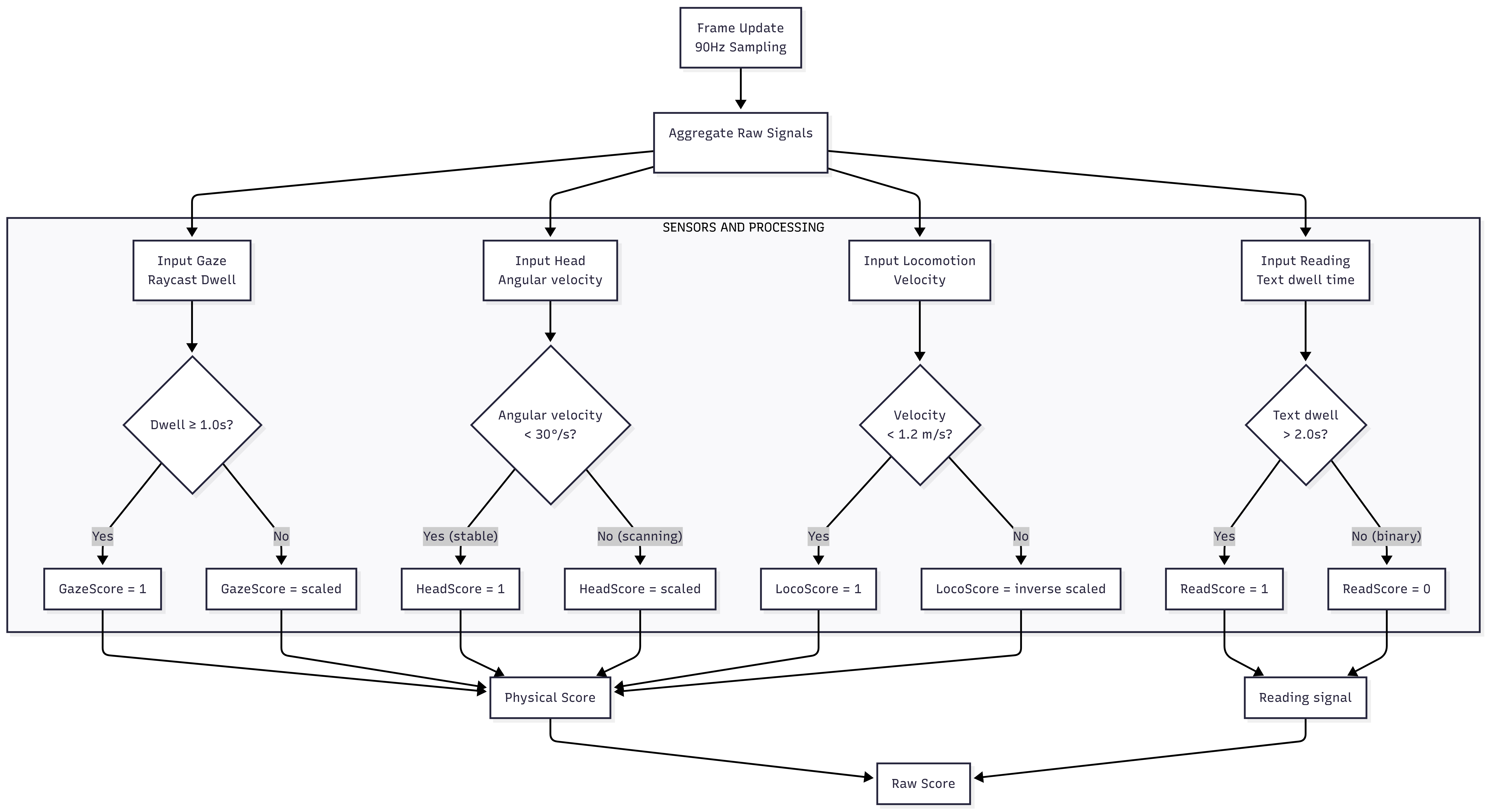}
  \caption{Signal Pre-processing and Sensor Fusion. Normalized behavioral signals are aggregated into a composite engagement score $E_{raw}$.}
  \label{fig:signal_fusion}
\end{figure}

Our engagement modeling architecture transforms raw sensor data into a high-level cognitive state estimation through a three-stage pipeline. In this work, we operationalize `engagement' specifically as a proxy for attentional stability and reading readiness, rather than a broader emotional or cognitive construct. Figure~\ref{fig:signal_fusion} summarizes the architecture, which is explained in subsequent sections.

\subsection{Signal Pre-processing}

Raw sensor data from the VR headset is sampled at 90\,Hz. To ensure robust classification, all inputs are normalized to a unit scale $[0,1]$.

\begin{itemize}
    \item \textbf{Head Stability ($S_{head}$):} Guided by research linking minimal head movement to sustained visual attention \cite{inproceedings}, we established a conservative threshold of \textbf{30\,$^\circ$/s} to differentiate between environmental scanning and focused fixation.
    \item \textbf{Gaze Dwell ($S_{gaze}$):} Prior work on gaze-based interaction suggests that dwell durations in the range of one second effectively distinguish intentional focus from transient glances \cite{10.1145/3607822.3614513}. Following this guidance, we adopt a \textbf{1.0-second dwell threshold} for gaze raycast stability in our system.
    \item \textbf{Locomotion Velocity ($S_{loco}$):} Research on VR locomotion indicates that users instinctively slow down when their cognitive load or engagement increases \cite{10.1145/3343036.3343119}. Based on typical VR walking speeds, we establish \textbf{1.2\,m/s} (approximating standard real-world walking speed) as a normalization baseline for comfortable locomotion; velocities exceeding this limit are inversely mapped to the engagement score.
\end{itemize}

\subsection{Sensor Fusion and Weighting}

Normalized signals are aggregated into a composite engagement score ($E_{raw}$) using a weighted fusion strategy ($W_{phys}=0.75$, $W_{read}=0.25$):
\begin{equation}
  E_{raw} = (W_{phys} \times S_{phys}) + (W_{read} \times S_{read})
\end{equation}

\begin{itemize}
    \item \textbf{Physical Score ($S_{phys}$):} Weighted average of Head Stability (35\%), Gaze (30\%), and Locomotion (35\%).
    \item \textbf{Reading Context ($S_{read}$):} Binary signal triggered when the user dwells on text for $>2.0$\,s \cite{lee2025reading}.
\end{itemize}

To model affective inertia \cite{DMELLO2012145}, the fused score is temporally smoothed using linear interpolation ($\alpha=0.35$).

\subsection{Safety Gates}

To reduce false positives during high-entropy states, the system implements conservative logic gates motivated by prior findings that locomotion increases cognitive load and reduces attentional availability in VR \cite{10.1145/3343036.3343119}.

\begin{enumerate}
    \item \textbf{Locomotion Gate ($v > 1.2$\,m/s):} During sustained locomotion beyond typical walking speed, users are unlikely to maintain stable visual attention to dense exhibit text. Consistent with previous research on locomotion as demanding cognitive resources and reducing processing capacity, we conservatively cap engagement inference at a \textit{Neutral} state.
    \item \textbf{Run Gate ($v > 2.0$\,m/s):} During high-speed locomotion exceeding typical walking velocities, users are unlikely to sustain focused content consumption. Drawing on evidence of velocity-dependent cognitive demands during VR locomotion, we conservatively treat running behavior as incompatible with focused reading and force the state to \textit{Disengaged}.
\end{enumerate}

\begin{figure}[htbp]
  \centering
  \includegraphics[width=0.65\textwidth, trim=40 40 50 40, clip]{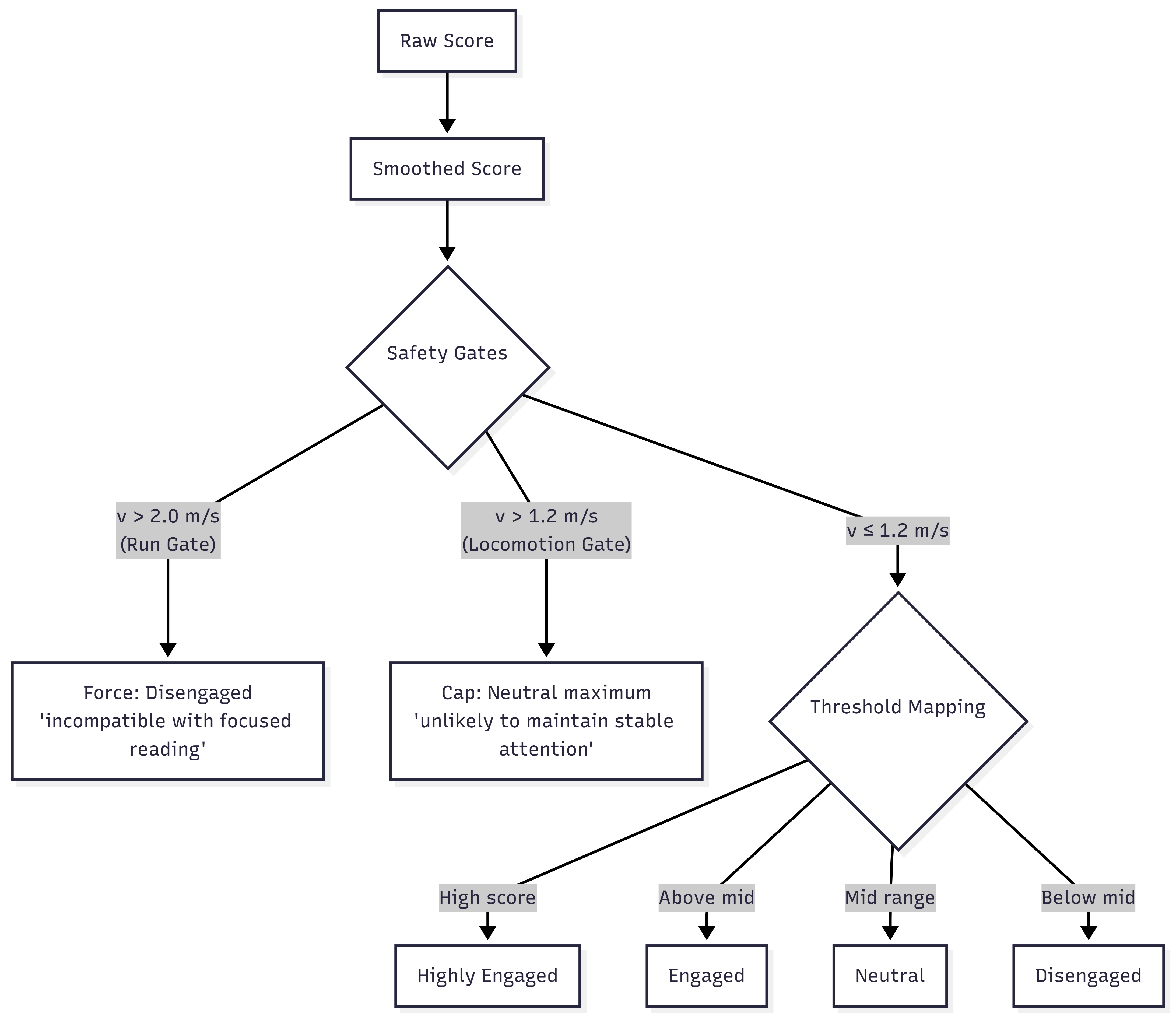}
  \caption{Temporal Smoothing and Safety Gates. The smoothed score passes through velocity-based safety gates before mapping to discrete engagement states.}
  \label{fig:safety_gates}
\end{figure}

Figure~\ref{fig:safety_gates} illustrates the temporal stabilization and safety gating applied to the raw engagement estimate. The continuously fused engagement score ($E_{\text{raw}}$) is first temporally smoothed using linear interpolation to model affective inertia and to prevent rapid oscillations between engagement states. The resulting smoothed score is then evaluated against velocity-based safety gates. During moderate locomotion ($v > 1.2\,\mathrm{m/s}$), engagement inference is conservatively capped at a \textit{Neutral} state, reflecting reduced attentional capacity while walking. During high-speed locomotion ($v > 2.0\,\mathrm{m/s}$), the system overrides the inferred score and forces a \textit{Disengaged} state. Only after temporal smoothing and safety gating is the engagement estimate mapped to discrete engagement states, which subsequently drive the adaptive content mechanism.

\subsection{Design Goals}

Our system supports real-time adaptive content in VR cultural heritage environments while remaining interpretable and deployable on consumer hardware. Four design goals guided the architecture:

\begin{enumerate}
   \item Adaptation relies solely on implicit behavioral signals from standard VR sensors, eliminating explicit input, calibration, or external instruments that could disrupt exploration.
   \item Engagement inference uses transparent, rule-based classification with defined thresholds rather than opaque machine learning, ensuring interpretability and reproducibility.
   \item Content adaptation operates under low-latency constraints compatible with VR frame rates, with sensing and inference executing in milliseconds to enable responsive adaptation.
   \item The system supports scalable content variation by leveraging generative language models, enabling adaptation across arbitrary exhibits without per-item scripting.
\end{enumerate}

\subsection{Participants}

Sixteen participants (8 per condition), aged 13--34 years, 15 men and 1 woman, were recruited through university mailing lists and social media. Table~\ref{tab:demographics} summarizes participant characteristics by condition. Mann--Whitney U tests confirmed that groups were balanced with respect to prior museum interest ($U = 30.0$, $p = .872$) and museum visit likelihood ($U = 28.5$, $p = .743$).

\begin{table}[htbp]
\centering
\small
\caption{Participant Characteristics by Condition}
\label{tab:demographics}
\begin{tabular}{lcc}
\toprule
\textbf{Characteristic} & \textbf{Adaptive (N=8)} & \textbf{Control (N=8)} \\
\midrule
\multicolumn{3}{l}{\textit{Prior VR Experience}} \\
Yes (once or twice) & 5 (62.5\%) & 4 (50.0\%) \\
No & 3 (37.5\%) & 4 (50.0\%) \\
\midrule
\multicolumn{3}{l}{\textit{Controller Gaming Frequency}} \\
Frequently (weekly/daily) & 4 (50.0\%) & 1 (12.5\%) \\
Occasionally & 2 (25.0\%) & 3 (37.5\%) \\
Rarely (monthly or less) & 2 (25.0\%) & 2 (25.0\%) \\
Never & 0 (0.0\%) & 2 (25.0\%) \\
\midrule
\multicolumn{3}{l}{\textit{Baseline Museum Attitudes, M (SD)}} \\
Interest in museums & 3.12 (1.25) & 3.25 (1.49) \\
Likelihood to visit & 3.75 (1.04) & 3.88 (1.36) \\
\bottomrule
\end{tabular}
\end{table}

\section{System Implementation}

The system follows a continuous closed-loop cycle: behavioral sensing, engagement inference, adaptive content selection, and user re-interaction. At each frame, multimodal behavioral signals from the VR headset are processed to estimate engagement state.

\begin{figure}[htbp]
  \centering
  \begin{subfigure}[b]{0.40\linewidth}
    \centering
    \includegraphics[width=\linewidth]{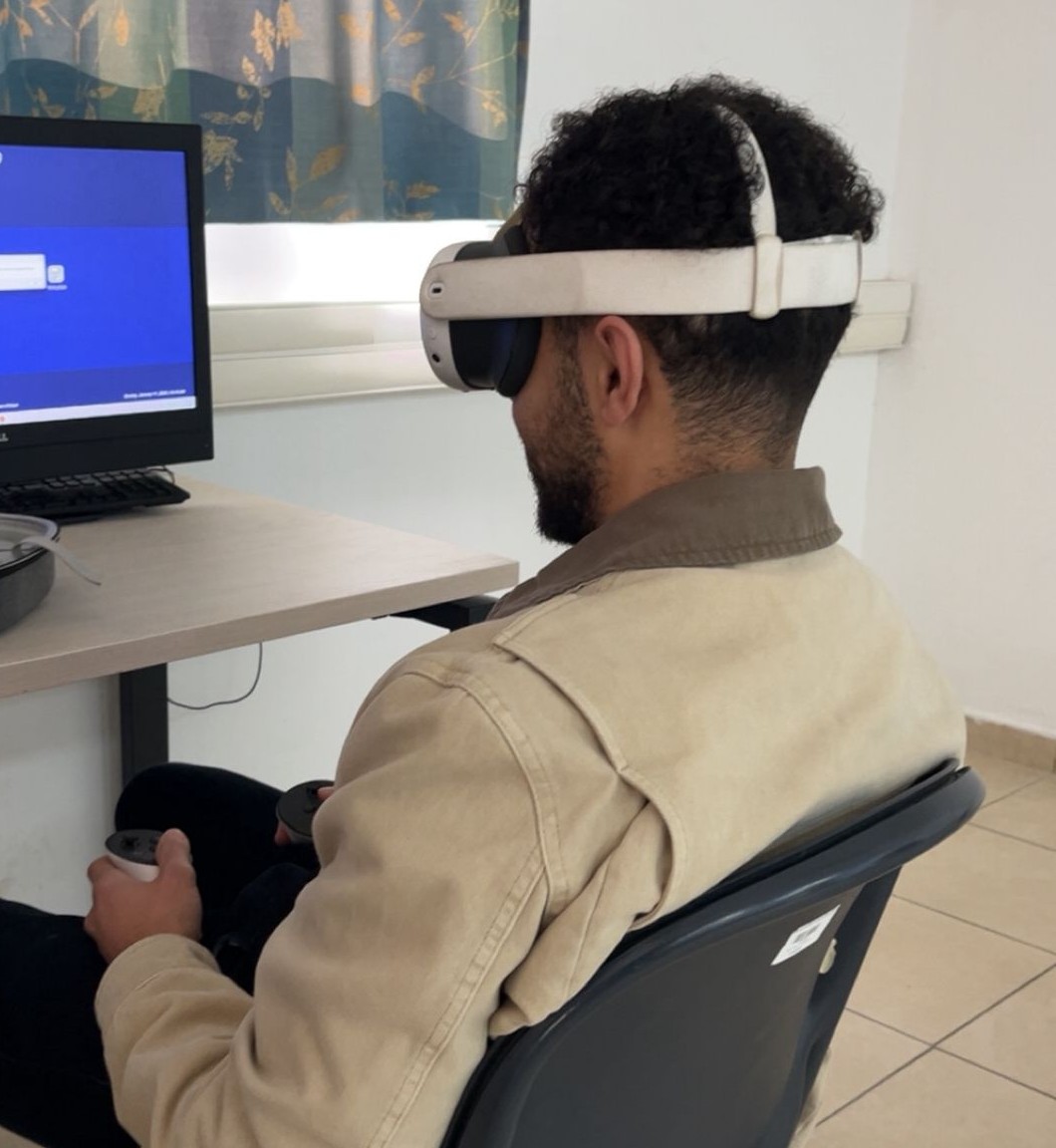}
    \caption{Participant during study}
    \label{fig:participant}
  \end{subfigure}
  \hfill
  \begin{subfigure}[b]{0.40\linewidth}
    \centering
    \includegraphics[width=\linewidth]{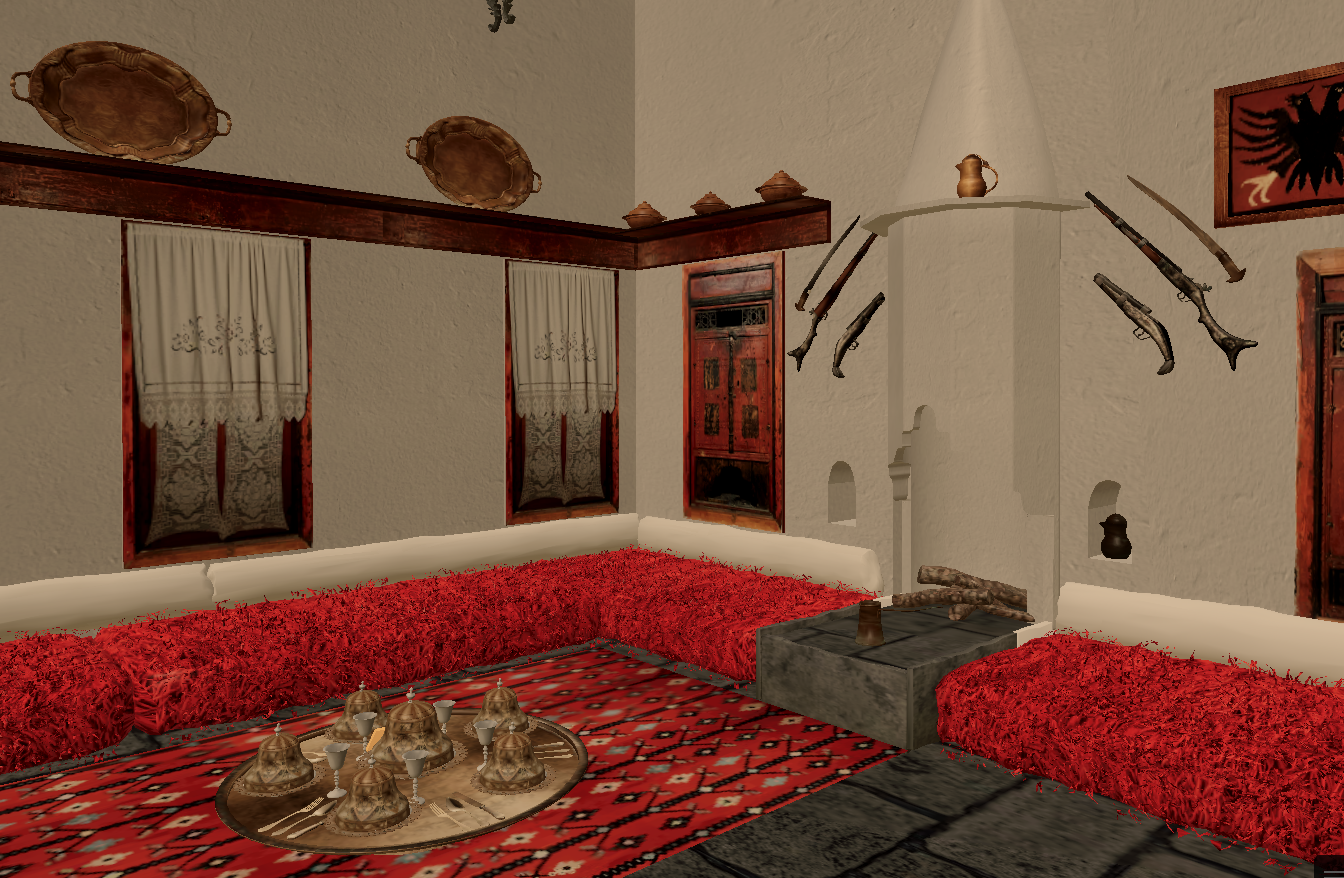}
    \caption{Virtual museum environment}
    \label{fig:museum}
  \end{subfigure}

  \vspace{6pt}

  \begin{subfigure}[b]{0.40\linewidth}
    \centering
    \includegraphics[width=\linewidth]{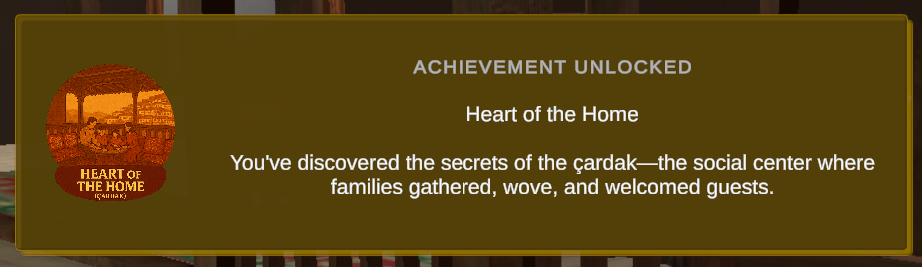}
    \caption{Achievements interface}
    \label{fig:achievements}
  \end{subfigure}
  \hfill
  \begin{subfigure}[b]{0.40\linewidth}
    \centering
    \includegraphics[width=\linewidth]{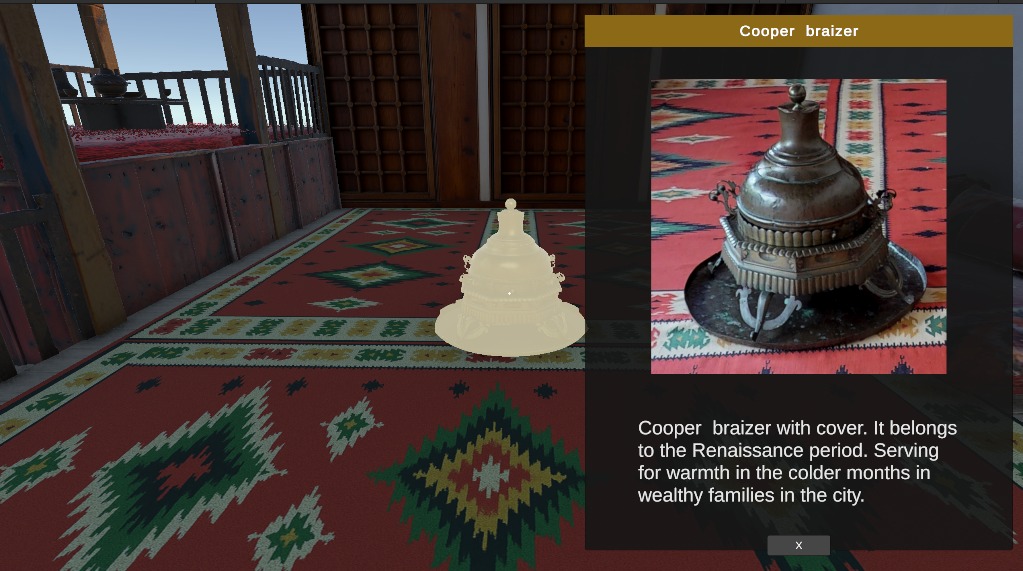}
    \caption{Object information panel}
    \label{fig:infopanel}
  \end{subfigure}

  \caption{System implementation overview: (a) participant wearing the Meta Quest 3 headset, (b) the virtual museum environment, (c) the achievements and gamification interface, and (d) an example object information panel showing adaptive content.}
  \label{fig:implementation}
\end{figure}

The system was developed using \textbf{Unity 6000.2.8f1} and deployed on a \textbf{Meta Quest 3}. All core interaction loops---behavioral sensing, signal aggregation, and engagement inference---execute locally on the Snapdragon XR2 Gen 2 processor for low-latency performance. Figure~\ref{fig:implementation} illustrates the study setup, virtual environment, and key interface components.

\subsection{Sensing and Inference}

Signal acquisition occurs within Unity's update and physics loops. Gaze is approximated via forward raycasting against exhibit colliders, while head kinematics and locomotion velocity derive from pose deltas of the \texttt{CenterEyeAnchor} and \texttt{CameraRig} transforms. The engagement inference engine applies temporal smoothing and safety gates before updating the global state machine.

\subsection{Asynchronous Content Pipeline}

Content adaptation is decoupled from the rendering thread to prevent stalls. Upon engagement state transitions, the system issues asynchronous requests to the LLM (GPT-4o). To ensure consistent output, we utilize a two-part prompting strategy: a System Instruction that emphasizes adherence to length constraints (e.g., ``You must stay within the requested word count''), and a dynamic User Prompt that adjusts the stylistic depth. As detailed in our supplementary material, this ranges from requesting a ``scholarly style with rich detail and anecdotes'' for Highly Engaged users, down to ``2--3 short bullet points under 10 words'' for Disengaged users. Generated explanations are cached using a composite key (exhibit ID + engagement level), enabling instant retrieval during revisits and reducing network overhead.

To concretely illustrate how engagement-driven prompting affects content presentation, Table~\ref{tab:content_adaptation} provides an example exhibit (a copper brazier) showing the contrast between static content and dynamically adapted outputs generated for different engagement states. These examples are illustrative and reflect the prompt strategies described above rather than fixed, hard-coded responses.

\begin{table}[htbp]
\centering
\caption{Comparison of Content Adaptation (Example Exhibit: Copper Brazier)}
\label{tab:content_adaptation}
\small
\setlength{\tabcolsep}{4pt}
\begin{tabularx}{\linewidth}{>{\raggedright\arraybackslash}p{2.6cm}
                            >{\raggedright\arraybackslash}p{3.6cm}
                            X
                            >{\raggedleft\arraybackslash}p{1.2cm}}
\toprule
\textbf{Engagement Level} &
\textbf{Prompt Strategy (Illustrative)} &
\textbf{Generated Content Example} &
\textbf{Words} \\
\midrule
Static / Neutral &
``Provide essential historical facts in a clear and concise manner.'' &
This copper brazier with cover dates to the Renaissance period. It served as a portable heater for wealthy families during colder months. The intricate floral engravings signify the owner's high social status. &
32 \\
\midrule
Disengaged &
``Format as 2--3 short bullet points, each under 10 words.'' &
\makecell[l]{Renaissance-era portable heater.\\Used by wealthy families.\\Floral engravings indicate high status.} &
13 \\
\midrule
Highly Engaged &
``Scholarly style with rich detail; include contextual anecdotes.'' &
This brazier is a masterwork of Renaissance metalworking. Beyond its function as a portable heater, the floral motifs on the lid were reserved for elite families, indicating use in formal guest spaces to impress visitors. &
46 \\
\bottomrule
\end{tabularx}
\end{table}

\section{Evaluation}
\label{sec:evaluation}

\subsection{Study Design}

We employed a between-subjects design with two experimental conditions ($N=16$):

\begin{itemize}
    \item \textbf{Adaptive (N=8):} The system continuously inferred user engagement from behavioral signals and dynamically adapted content depth in real time.
    \item \textbf{Control (N=8):} The system presented static, medium-length content for all exhibits without behavioral sensing or adaptation.
\end{itemize}

Both conditions used identical virtual environments, exhibit assets, interaction techniques, and navigation mechanics. The only difference between conditions was the presence of engagement-driven content adaptation.

\subsection{Procedure}

After providing informed consent, participants completed a brief demographic questionnaire and received standardized instructions on how to navigate and interact within the VR museum using a Meta Quest~3 headset. Participants were instructed to explore the museum freely, read exhibit descriptions as desired, and collect information cards distributed throughout the environment.

Each participant completed a single exploration session lasting up to 15 minutes. Following the VR experience, participants immediately completed post-experience questionnaires measuring usability, perceived presence, motivation to read, and interest in visiting the physical museum. Participants also provided optional qualitative feedback. The total session duration was approximately 20--30 minutes per participant.

\subsection{Measures}

\subsubsection{Objective Behavioral Measures}

The system automatically logged:
\begin{itemize}
    \item \textbf{Session Duration:} Total time spent in the VR experience.
    \item \textbf{Reading Behavior:} Number of reading events, total reading view time, and estimated number of words exposed.
    \item \textbf{Interaction Outcomes:} Number of cards collected (maximum 18) and badges unlocked.
    \item \textbf{Engagement States:} Distribution of time across engagement classifications.
\end{itemize}

\subsubsection{Subjective Measures}

Usability was assessed using the System Usability Scale (SUS), producing scores from 0 to 100. Additional Likert-scale items (1--5) measured perceived presence, motivation to read exhibit text, navigation ease, and content appropriateness. Participants also reported whether the VR experience increased their interest in visiting the physical museum.

\subsection{Results}

Given the modest sample size ($N=16$), Mann--Whitney U tests were used with Cohen's $d$ effect sizes reported. Table~\ref{tab:behavioral} presents behavioral outcomes and Table~\ref{tab:engagement} presents engagement state distribution. Overall system usability was high, with a mean SUS score of 84.3 across both conditions.

\begin{table}[htbp]
\centering
\small
\caption{Behavioral Outcomes by Condition}
\label{tab:behavioral}
\begin{tabular}{lcccc}
\toprule
\textbf{Metric} & \textbf{Adaptive} & \textbf{Control} & \textbf{p} & \textbf{d} \\
 & M (SD) & M (SD) & & \\
\midrule
Session Duration (s) & 889.4 (422.0) & 451.4 (157.3) & .003** & 1.38 \\
Reading View Time (s) & 473.4 (294.5) & 164.1 (121.1) & .015* & 1.37 \\
Reading Events & 67.2 (19.9) & 50.1 (34.7) & .050* & 0.61 \\
Words Exposed & 1515.5 (616.8) & 745.1 (511.3) & .007** & 1.36 \\
Reading Motivation (1--5) & 4.00 (0.76) & 3.25 (0.89) & .083 & 0.91 \\
Cards Collected & 8.6 (5.4) & 4.8 (3.7) & .140 & 0.84 \\
\bottomrule
\multicolumn{5}{l}{\small *$p<.05$, **$p<.01$; Mann--Whitney U tests}
\end{tabular}
\end{table}

\begin{table}[htbp]
\centering
\small
\caption{Engagement State Distribution (\% of Session Time)}
\label{tab:engagement}
\begin{tabular}{lccccc}
\toprule
\textbf{Condition} & \textbf{Highly Eng.} & \textbf{Engaged} & \textbf{Neutral} & \textbf{Disengaged} & \textbf{Highly Dis.} \\
\midrule
Adaptive & 24.3\% & 47.4\% & 22.2\% & 5.9\% & 0.3\% \\
Control & 13.6\% & 48.0\% & 28.4\% & 9.6\% & 0.3\% \\
\bottomrule
\end{tabular}
\end{table}

\paragraph{Subjective Measures.}
Perceived presence trended higher in the Adaptive condition (Adaptive: $M = 4.12$, $SD = 0.35$; Control: $M = 3.50$, $SD = 1.20$; $U = 42.5$, $p = .201$), suggesting that adaptive text generation did not disrupt immersion and may have enhanced it. Interest in visiting the physical museum also trended higher in the Adaptive condition (Adaptive: $M = 3.75$, $SD = 0.89$; Control: $M = 3.25$, $SD = 1.16$; $U = 40.0$, $p = .414$), though neither difference reached statistical significance given the sample size. Due to the exploratory nature of the study, these self-report findings should be interpreted cautiously.

\subsection{Discussion}

Our preliminary results suggest that LLM-driven content pacing, guided by implicit multimodal sensing, can increase behavioral indicators of engagement in immersive cultural heritage experiences.

\paragraph{RQ1: To what extent can implicit behavioral signals support real-time attention-aware adaptation?}
The results indicate that the selected implicit behavioral signals were sufficient to support stable, real-time adaptation of content depth without disrupting the user experience. While these signals do not provide direct access to users' internal cognitive states, their combination enabled the system to consistently differentiate between scanning behavior and sustained reading readiness, allowing timely transitions between engagement states. The absence of usability degradation (SUS $>$ 84) and the smooth operation of the adaptive pipeline suggest that attention-aware adaptation based on implicit signals is feasible within the latency and interaction constraints of consumer VR systems. These findings position implicit multimodal sensing as a practical foundation for attention-aware content pacing rather than a definitive measure of cognitive engagement.

\paragraph{RQ2: How does adaptation influence exploration behavior, content consumption, and motivation to engage?}
The adaptive condition demonstrated higher positive engagement, with 71.7\% of session time in combined \textit{Highly Engaged} and \textit{Engaged} states compared to 61.6\% in the control condition. Notably, \textit{Highly Engaged} time nearly doubled (24.3\% vs.\ 13.6\%), while combined negative engagement (\textit{Disengaged} + \textit{Highly Disengaged}) decreased by 37\% (6.2\% vs.\ 9.9\%). The adaptive condition also reduced neutral or passive time from 28.4\% to 22.2\%, suggesting that users spent more time actively engaged rather than passively browsing.

These patterns indicate that the LLM-mediated adaptive scaffolding helped maintain flow by providing appropriately challenging content, reducing cognitive friction and disengagement while keeping users within productive learning zones \cite{lee2025reading,DMELLO2012145}. This observation aligns with prior work showing that adaptive and affect-aware systems can positively influence engagement during complex learning and exploratory tasks without disrupting user experience \cite{Yildirim17012025,khan2025museum}.

\paragraph{Trust, Transparency, and Multimodal Interaction.}
While participants were not informed texts were AI-generated, none questioned content authenticity, though this raises ethical questions about transparency in AI-mediated cultural experiences. Scalability beyond our constrained domain would require safeguards against hallucination, potentially through retrieval-augmented generation grounding LLM outputs in curated knowledge bases \cite{cho2026mobile}. Our multimodal sensing approach demonstrates that effective human-AI collaboration need not rely on explicit commands \cite{clay2023}. The success of purely implicit signals suggests embodied behavior serves as a rich communication modality, enabling proactive AI support without breaking presence \cite{10.1145/3343036.3343119}. However, future work should investigate user awareness of AI adaptation and desires for explicit control.

\subsection{Limitations}

The modest sample size ($N=16$) limits statistical power for detecting non-significant effects, even when effect sizes are moderate to large ($d \approx 0.87$).

Our engagement inference relies on a fixed 4-second rolling window to smooth signal noise. Informal pilot observations suggested that shorter windows ($<2$\,s) introduced instability due to transient glances, while longer windows ($>6$\,s) led to perceived system responsiveness lag, highlighting the need for future personalization of temporal thresholds.

Additionally, metrics such as words exposed are intrinsically influenced by the intervention itself, as the adaptive system presents longer text to users inferred as more engaged; this measure therefore reflects both system behavior and user choices rather than engagement alone.

We did not assess knowledge retention or validate engagement classification against external ground-truth measures, which remains an important direction for future multimodal engagement modeling in immersive museum experiences \cite{sawyer2020,clay2023}. The 15-minute exploration sessions capture initial interaction patterns but do not reflect long-term engagement or learning effects. Finally, privacy implications of continuous behavioral sensing and AI-mediated adaptation warrant further investigation \cite{tang2025}.

\section{Conclusions and Future Work}

This work introduces a closed-loop framework in which a Large Language Model responds to implicit multimodal behavioral signals in a VR cultural heritage environment, leading to substantially increased exploration time, reading engagement, and word exposure while maintaining high usability. Our findings suggest that AI-integrated XR systems can leverage multimodal behavioral data as practical signals for attention-aware adaptation, allowing the LLM to act as a dynamic mediator that adapts authored content to user capacity by balancing curated historical accuracy with personalized pacing. Future work should investigate long-term effects, explicit validation of engagement proxies, and the integration of multimodal foundation models that enable agents to suggest exhibits, answer questions, or provide collaborative annotations based on observed user interests.

\section*{Acknowledgments}

We thank all participants for their time and valuable feedback. The virtual reconstruction was informed by publicly available materials and virtual tour content provided on the official website of the Berat National Museum.

\section*{Declaration on Generative AI}

This work integrates the OpenAI GPT-4o API as part of the adaptive content generation pipeline described in the system implementation. GPT-4o was used to dynamically generate exhibit descriptions at varying levels of complexity during runtime. AI-assisted tools were also used for minor grammar and language refinement during manuscript preparation. The authors reviewed and edited all content and take full responsibility for the final manuscript.

\bibliographystyle{plainnat}
\bibliography{references}

\end{document}